\documentclass[useAMS,usenatbib]{mn2e}
\usepackage{graphicx}
\usepackage{epstopdf}
\usepackage{amsmath}
\title[Detection of the fine structure of the pulsar J0953+0755 radio emission in the  decametre wave range]{Detection of the fine structure of the pulsar J0953+0755 radio emission in the  decametre wave range}
\author[Ulyanov et al.]
{O. M. ~Ulyanov,$^1$\thanks{E-mail:
		oulyanov@rian.kharkov.ua}
	A. O. ~Skoryk,$^1$ A. I. ~Shevtsova$^1$, M. S. ~Plakhov$^1$
	\newauthor and O. O. Ulyanova$^1$
	\\
	$^1$ Department of Astrophysics, Institute of Radio Astronomy of NAS of Ukraine, 
	Krasnoznamennaya str. 4, Kharkov 61002, Ukraine}

\date{Released 2014 Xxxxx XX}

\pagerange{\pageref{firstpage}--\pageref{lastpage}} \pubyear{2014}

\begin{document}
	
	\label{firstpage}
	
	\maketitle
	
	\begin{abstract}
		In this paper the anomalous intense pulse of the PSR J0953+0755 was studied in decametre wavelength range. For this pulse two scales of fine structure were discovered. The long-scale structure consists of four components, where the visible dispersion measures of even and odd components are different. The obtained time-scale of the short fine structure is 1 ms. The difference in visible dispersion measure can be caused by propagation of two normal modes of the pulsar radiation and irregularities of electron concentration in the space near the neutron star like upper layers of magnetosphere and pulsar wind. 
	\end{abstract}
	
	\begin{keywords}
		magnetic fields -- plasmas -- polarization -- pulsars: general, pulsars: individual: J0953+0755
	\end{keywords}
	
	\section{Introduction}
	
	An individual pulsar pulse may show an intensity modulation at scales less than the main pulse duration. This phenomenon is called fine structure of pulsar radiation.  First microsecond intensity variations were observed by \cite{1971ApJ...169..487H}. 
	The main parameter of the fine structure is coherence time called characteristic time-scale. The time-scale is frequency dependent and ranges from milliseconds at decametre observation wavelengths \citep{2012ARep...56..417U, 2013IAUS..Ulyanov}, microseconds at metre wavelengths \citep{1981SvA....25..442S, 1994A&A...286..807S, 2000ASPC..202..181S, 2001ARep...45..294S} up to nanoseconds at centimetre  wavelengths \citep{2003Natur.422..141H, 2007ApJ...670..693H}. 
	
	The existing classification of the fine structure includes nanostructure, microstructure and subpulse structure. In this paper we use general term of fine structure, that may include different scales of coherence at the same time. 
	
	Subpulses have been studied since 1968 \citep{1968Natur.220..231D} and are believed to be basic structure units of pulsar pulses. Many authors, as \cite{1986ApJ...301..901R}, \cite{2003A&A...407..315G}, \cite{2004IAUS..218..321G}, \cite{2013mnras...Rankin}, say that subpulses may explain macro parameters of pulsar radiation plasma dynamics or magnetic field geometry. Also, subpulses may include shorter pulses so that two scales (long and short) are observed at the same time \citep{1981SvA....25..442S, 1987SvA....31..529P}. At decametre wavelengths  long-scale is usually milliseconds and short-scale is submilliseconds. Comparatively short-structure has duration of several microseconds in decimetre and centimeter ranges \citep{1981SvA....25..442S}. It is supposed that scattering limits the lower range of fine structure duration \citep{1981SvA....25..442S, 1987SvA....31..529P}. Sometimes intensity of the fine structure is quasi-periodic \citep{1976ApJ...208L..43B, 1981IAUS...95..199B, 1983SvA....27..169S}. 
	
	The microstructure has been investigated mainly in decimeter and meter wavelength ranges. In decametre range only few studies of micropulses were made, whereas more attention was paid to subpulse structure.  Some results of microstructure studies were presented in \cite{1984AZh....61..343N}, where authors detected a millisecond fine structure of PSR B0809+74 radiation at 25 MHz using the UTR-2 radio telescope. The time-scale of this structure was 2 -- 4 ms. 
	
	It is hard to detect an individual pulsar pulse at low frequencies as it is scattered by propagation medium. However, a phenomenon of anomalous intense pulse (AIP) that was found at decametre \citep{UlyanovZaharenko2006, UlyanovDeshpande2007, 2008ARep...52..917U} and metre \citep{1998ARep...42..241M, 2003AstL...29...91E} observational wavelengths makes a fine structure of pulsar radiation be possible to explore. Having intensity hundred times larger than an average pulse, AIPs have enough signal-to-noise ratio (SNR) to be detectable. This gives opportunities to study processes at the short time-scales in this frequency range. However, the probability of AIP registration at decametre wavelengths is only 1 -- 2 per cent \citep{UlyanovZaharenko2006}. The AIPs of the PSR J0953+0755 were detected by \cite{2012ARep...56..417U}. 
	
	Currently there are two most complete models of the fine structure origin. The first one relies on broadbandness property and explains the fine structure as a result of radial modulation such as neutron star vibrations or polar cap periodical sparking \citep{1981IAUS...95..191B, 1981IAUS...95..199B}. Another model \citep{2004A&A...417L..29P} claims that fine structure is caused by induced scattering. Still there is no clear understanding of the fine structure phenomenon. 
	
	The aim of this paper is to explain at least some parameters of pulsar radiation fine structure by propagation effects in magnetoactive inhomogeneous plasma \citep{2013IAUS..Ulyanov}. One of the main effects under consideration is time dispersion delay caused by free charges (electrons and positrons in the pulsar magnetosphere, electrons in the interstellar medium) and irregularities of the plasma along a line of sight. The dispersion delay in the interstellar medium (ISM) is well explained by formulas \citep{1971ApJ...169..487H}. Besides, there are scattering, the Faraday effect, refraction and birefringence phenomenon. 
	
	The decametre waverange is promising to study fine effects of propagation medium and probe a pulsar magnetosphere in depth. At low frequencies all propagation effects are the most prominent. At the same time, the widest relative frequency band $\Delta f / f_c$ (where $f_c$ is the central frequency) provides us with an opportunity to observe these effects in a wide range using a single radio telescope under the same conditions.
	
	We follow the approach of the work \citet{2003A&A...408.1057P, 2004A&A...417L..29P, 2006MNRAS.366.1539P}, and consider a pulsar magnetosphere to be one of the propagation medium part. A pulsar magnetosphere is highly magnetized. In the strong magnetic field normal modes of electromagnetic waves are linearly and elliptically polarized in a wide range of angles between a wave vector $\mathbf{k}(\omega)$ and a magnetic induction vector $\mathbf{B}$ \citep{1977ewcp.book.....Z, Zheleznyakov1997, 2013BaltA..22...53U, UlyanovShevtsova2013, UlyanovShevtsova2014}. Moreover, refractive coefficients of ordinary and extraordinary waves are different and two wave modes pass different electrical distance to an observer. That would mean a spatial dispersion of the normal wave modes in a pulsar magnetosphere similarly to frequency dispersion. Hence we cannot use one single DM to transform a signal to a pulsar reference frame as it is common at the present time (at frequencies above 100 MHz).
	
	A magnetic field intensity falls rapidly from a pulsar surface to radius of the light cylinder. The dipole term of the magnetic field drops dramatically as $\mathbf{B} \sim R_{sur}^3/r^3$, where $r$ is a distance from a pulsar centre, $R_{sur}$ is a neutron star radius. Typical magnetic field in the ISM is 0.5 -- 1 ${\mu} \mbox{G}$, average electron concentration in the ISM is $ <N_{ISM}> \approx 0.03 \mbox{~cm}^{-3}$. It means that elliptically polarized waves coming from the pulsar magnetosphere can be represented in the weak anisotropic ISM  as normal modes with right and left circular polarization in a wide range of angles between $\mathbf{k}(\omega)$ and $\mathbf{B}$ \citep{2013BaltA..22...53U}. In this case we can use quasi-longitudinal propagation approach up to low frequency $ f \sim$  20 MHz \citep{UlyanovShevtsova2013, UlyanovShevtsova2014}. 
	
	In this paper we analyse the pulsar signal that propagates through upper magnetosphere layers. We assume that a visible radiation region is surface at the polarization-limiting radius. Above this surface polarization properties of pulsar radiation do not change \citep{1970pewp.book.....G, 1977ewcp.book.....Z, Zheleznyakov1997, 2001A&A...378..883P, 2003A&A...408.1057P, 2006MNRAS.366.1539P, 2006MNRAS.368.1764P}. It is believed that the emission region is far from a neutron star surface. However, some authors suggested that radio emission may originate at centimeter altitudes near a star surface \citep{2007JETPL..85..267K, 2013Ap&SS.345..169K}. 
	
	In the current paper we choose the PSR J0953+0755 to analyse since it is the nearest pulsar. It means that it has the lowest scattering measure, DM and rotation measure (RM). And this pulsar is one of the best candidates to study fine structure effects at low frequencies. We analyse one particular AIP of the PSR J0953+0755 using spectral and correlation analysis. We obtain the intensity fine structure of its radiation with different characteristic time-scales, we show the differences in DM values of the pulse components and estimate some numerical parameter of the PSR J0953+0755 magnetosphere. Using the shortest scales of fine structure as a probing pulses provides us with opportunity to explore spatial properties of the pulsar magnetosphere and the polarization-limiting radius.
	
	\section{Observations, data processing and results}
	
	\begin{figure*}
		\includegraphics[width=0.85\textwidth]{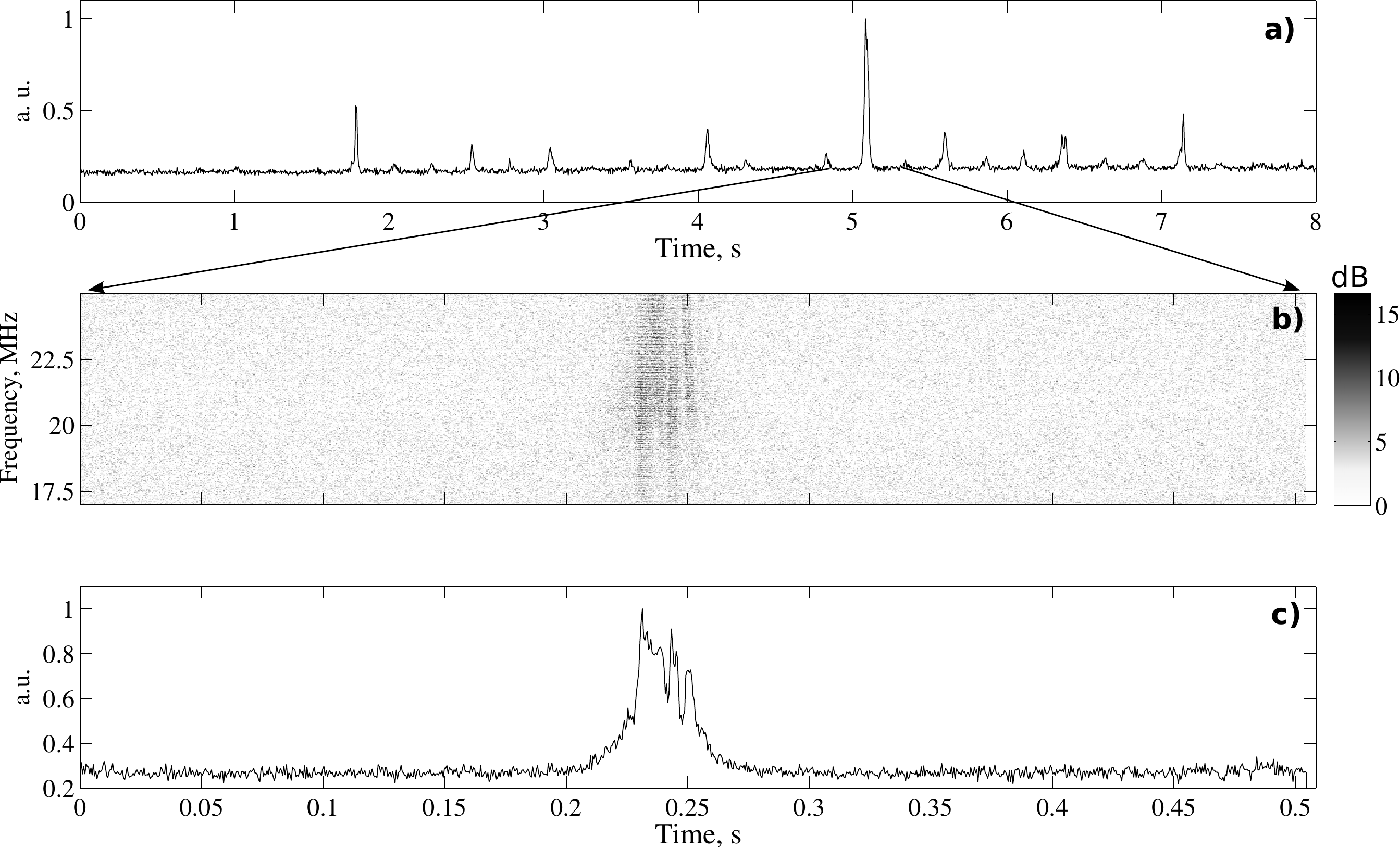} 
		\caption{a) The sequence of the PSR J0953+0755 AIPs; b) the dynamic spectrum of the most strong AIP with high time resolution ($\Delta \tau_{DS} = 62 ~\mu \mbox{s}$, $\Delta f_{DS} = 32$ kHz); c) the temporal profile of the most strong AIP. $\mbox{DM} = 2.972~\mbox{pc}~\mbox{cm}^{-3}$, frequency range 17 -- 25 MHz. }
	\end{figure*}
	
	Observations of the pulsar J0953+0755 were made on 2013 February 22 using the Ukrainian T-shaped Radio telescope (UTR-2) \citep{1978MenSodin, 2005KFNTS...5...90Z}. The linear size of north-south arm is 1854 m and the west-east arm is 927 m. At the central observation frequency 18 MHz the UTR-2 effective area is close to $150 \cdot 10^3 \cdot \cos(z) ~\mbox{m}^2$, where $z$ is a zenith angle. The \mbox{UTR-2} consists of 2040 broad-band Nadenenko dipoles with one linear polarization. The main axes of all dipoles are parallel to the west-east direction. UTR-2 covers a frequency range from 8 MHz up to 33 MHz. Signals from PSR J0953+0755 were registered on the digital receiver DSPZ \citep{2010A&A...510A..16R} in the Wave Form mode \citep{2007RRPRA..12..233Z} using 16-bit analogue-to-digital converter. Sampling rate was 66 MHz, the band-pass filter had 18 -- 30 MHz pass band (the cut-off frequencies were for 3 dB level). UTR-2 operated in source tracking mode and the interval of its antenna pattern switching was 2 minutes. The widths of the main lobe were $0^{\circ}.5$ (in plane of declination) and $15^{\circ}$ (in plane of right ascension) near the local meridian. The intensity modulation of radio emission due to the pointing errors of the UTR-2 radiation pattern while a source was moving across the main lobe was minimal and did not exceed 1.5 dB. All the telescope parameters are presented for frequency $ f = 25$ MHz.  
	
	During the observations a sequence of AIPs of the PSR J0953+0755 was detected (Fig. 1 a). 
	
	The main attention was paid on the most strong pulse from this sequence. We used a classical approach of spectral and correlation analysis to detect fine structure. An auto correlation function of intensity (ACF) may show features that point to fine structure presence with characteristic time-scales \citep{1975ApJ...197..185R}. This data processing technique we have used can also be applied to studies of transients, which are aperiodic sources \citep{2006Natur.439..817M}.
	
	First of all, we need to compensate time delays due to dispersion. Several DM values were offered in literature (see for example \cite{2013MNRAS.431.3624Z, 2015AJ....149...65T}). Taking a target value of  DM = 2.972 $~\mbox{pc~cm}^{-3}$ from \citep{2013MNRAS.431.3624Z} we removed the dispersion delay from the pulsar signal using coherent technique \citep{1971ApJ...169..487H, 1975MComP..14...55H}. By taking this we obtained high resolution temporal profile of the AIP sequence (time resolution was $1/\Delta f \approx 83$ ns, $\Delta f = (30-18) = 12$ MHz). Then we constructed a dynamic spectrum from the already compensated temporal profile of the pulsar data. The time and frequency resolutions of the dynamic spectrum were $\Delta \tau_{DS} = 62 ~\mu \mbox{s}$ and $\Delta f_{DS} = 32$ kHz respectively. The profile with high time resolution and corresponding dynamic spectrum in the frequency range 17--25 MHz can be seen in Fig. 1.  
	
	The analysis of the temporal profile and the dynamic spectrum of the AIP showed presence of at least four components in the 20 -- 28 MHz range. At the frequencies higher than 28 MHz the intensity of detected signal fell. At frequencies 28 -- 30 MHz the UTR-2 effective area drops dramatically as the antenna array becomes rarer with increasing frequency and side lobes appear. This leads to decrease in the telescope sensitivity and hinders the AIP detection above 28 MHz frequency. On the other hand, at the frequencies lower than 20 MHz the Galactic background brightness temperature increases \citep{Sidorchuk2008} and AIP flux density falls \citep{2012ARep...56..417U} at the same time, resulting in suppression of the AIP intensity by the galactic background intensity. 
	
	Using coherent method of dispersion removal we obtained four visible DM values for each component of the AIP. This approach is used to find the maximum intensity of a component versus DM \citep{1972ApJ...177L..11H}. The obtained correct dispersion measures for each component were $\mbox{DM}_1 = 2.9720$, $\mbox{DM}_2 = 2.9731$, $\mbox{DM}_3 = 2.9720$, $\mbox{DM}_4 = 2.9729 ~\mbox{pc}~\mbox{cm}^{-3}$.
	
	Errors of the DM estimation when using the coherent dispersion compensation are caused by instrumental and physical parameters of the propagation medium. Instrumental errors are associated with the delay time, which is inversely proportional to a filter bandwidth. In our case the filter band was 18 -- 30 MHz which corresponded to the accuracy $\delta \mbox{DM}_{hw} \approx 2 \cdot 10^{-8} ~\mbox{pc}~\mbox{cm}^{-3}$. Physical resolution is mainly restricted by scattering. The characteristic DM fluctuation at time-scales 1 ms (the preliminary scattering time of the PSR J0953+0755 at 25 MHz taken from \citep{2012ARep...56..417U}) is $\delta DM_{med} \approx 1.2 \cdot 10^{-4} ~\mbox{pc}~\mbox{cm}^{-3}$. Hence, resulting accuracy of DM estimation is $\delta DM_{tot} \approx 1.2 \cdot 10^{-4} ~\mbox{pc}~\mbox{cm}^{-3}$. This accuracy is sufficient to detect fine variations of a visible dispersion measure inside the individual pulsar pulse.
	
	One can see in Fig.1 (b), Fig.3 (a) that the first and the third components of the AIP have almost the same visible value of DM. Similarly for the second and the fourth components (see Fig.3 (b)). Evidently, we can split four components in two pairs of odd and even components. The visible dispersion measures of two pairs are different from each other and have the average value $\mbox{DM}_{odd} = (DM_1+ DM_3)/2 = 2.9720 \pm 1.2 \cdot 10^{-4} ~\mbox{pc}~\mbox{cm}^{-3}$ and $\mbox{DM}_{even} = (DM_2 + DM_4)/2 = 2.9730 \pm 1.2 \cdot 10^{-4} ~\mbox{pc}~\mbox{cm}^{-3}$. To exclude the impact of component pairs on each other we studied the pairs independently by segregating the components of the same pair. Segregation was done by applying a window in a time domain. The window had double-humped shape. The width of each “hump” was approximately 5.5 ms and the distance between “humps” was 8 ms and 12 ms for different component pairs. This method increased the SNR of the data by suppressing components with $\mbox{"wrong"}$ DM.
	
	The next step of data processing consists in obtaining two temporal profiles of the AIP with different time resolutions (see Fig. 2). The algorithm for the odd pair is shown below. The other pair is processed in the same way, but using another DM while dispersion compensating. 
	
	We obtained the first temporal profile of the AIP of PSR J0953+0755 after dispersion removing by coherent method with $\mbox{DM}_{odd} = 2.9720 ~\mbox{pc}~\mbox{cm}^{-3}$ (Fig. 2 a). The frequencies of registration were 18 -- 30 MHz and the time resolution of the profile after smoothing was $\Delta \tau_{coh} \approx 4 \mu \mbox{s}$. In our conception this profile may contain a fine structure of time-scale of the same order as the scattering time-scale of this pulsar at 25 MHz which is 1 ms (taken from \citet{2012ARep...56..417U}).
	
	Another low resolution profile of the same pulse data was obtained by using the post detection method of dispersion removal. The dynamic spectrum of the main AIP was summed over all frequencies, so the obtained average profile has $\Delta \tau_{pd} \approx$ 4 ms time resolution. But then the time resolution was linearly interpolated to the $\Delta \tau_{pd} ' = \Delta \tau_{coh} \approx 4  ~\mu \mbox{s}$ to match with the first profile (Fig. 2 b). As a result we obtained two pulse profiles of the same length (number of time samples ) but with different physical correlation scale. While the fine structure may be contained in the high resolution pulse, it is completely smoothed in the low resolution pulse.
	
	\begin{figure}
		\includegraphics[width=0.5\textwidth]{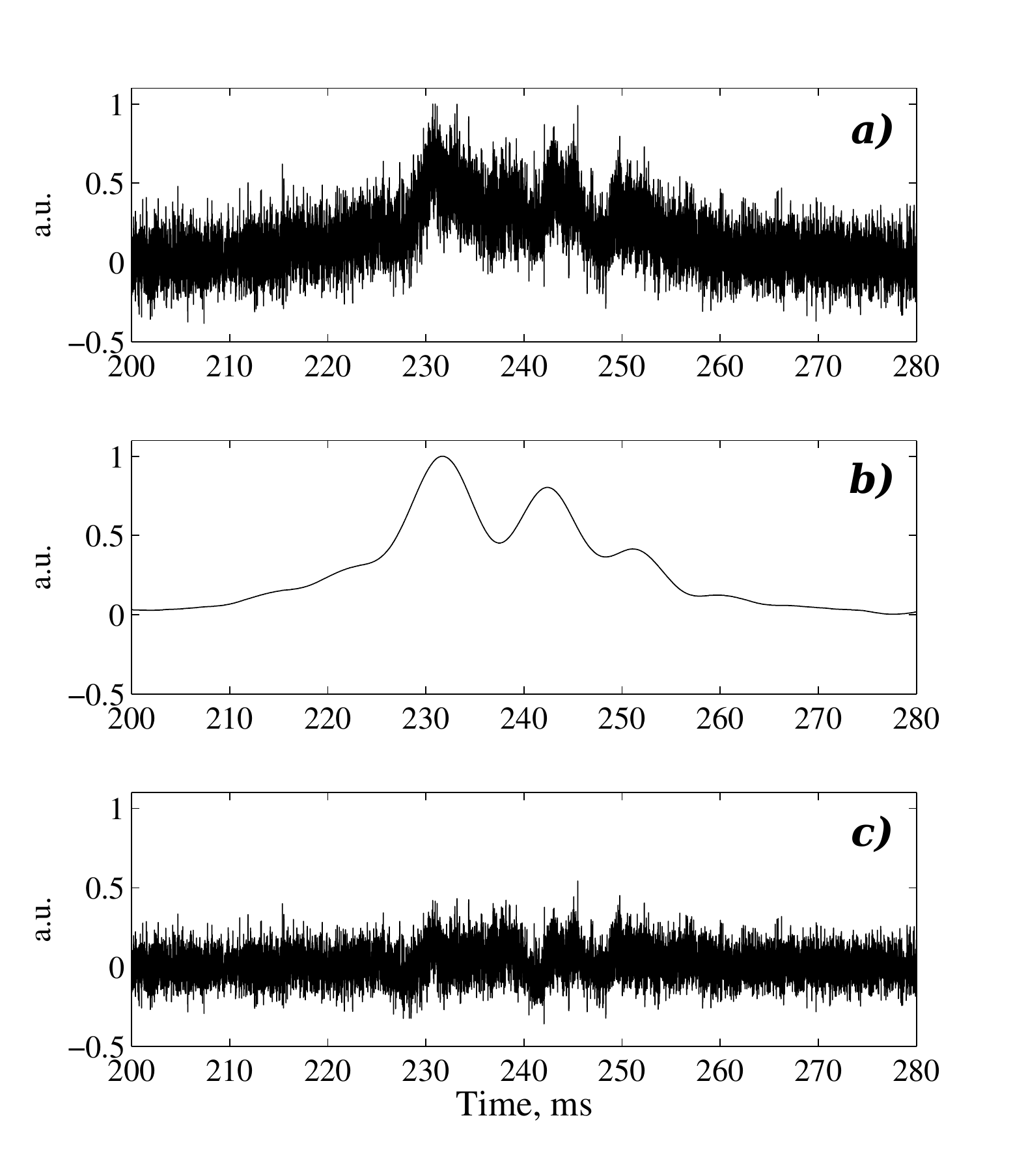} 
		\caption{The temporal profiles of the same AIP of PSR J0953+0755 with a) $4 ~\mu \mbox{s}$ time resolution; b) 4 ms time resolution. c) The differential profile of the upper profiles. $\mbox{DM}_{odd} = 2.972 ~\mbox{pc}~\mbox{cm}^{-3} $.}
	\end{figure}
	
	We normalized the profiles to make the total energy in the pulse windows approximately equal and then we subtracted intensities of one pulse from another (with a coefficient 0.8 for the low resolved profile). The differential profile is shown in the Fig. 2 c. Thus, we have increased the contrast of a fine structure contained in the high resolution pulse. The differential signal was analysed by spectral and correlation analysis. The other pair of components was analysed in a similar way taking the dispersion measure $2.9730 ~\mbox{pc}~\mbox{cm}^{-3}$ and its appropriate temporal intervals of the window.
	
	This approach of separating the component pairs of different visible DM from each other using double-humped shaped window and taking difference of two profiles with high and low physical resolution enables us to detect the fine structure of the PSR J0953+0755 radiation at decametre wave lengths. The results are shown below.
	
	\section{The results interpretation}
	
	\begin{figure*} 
		\includegraphics[width=1\textwidth]{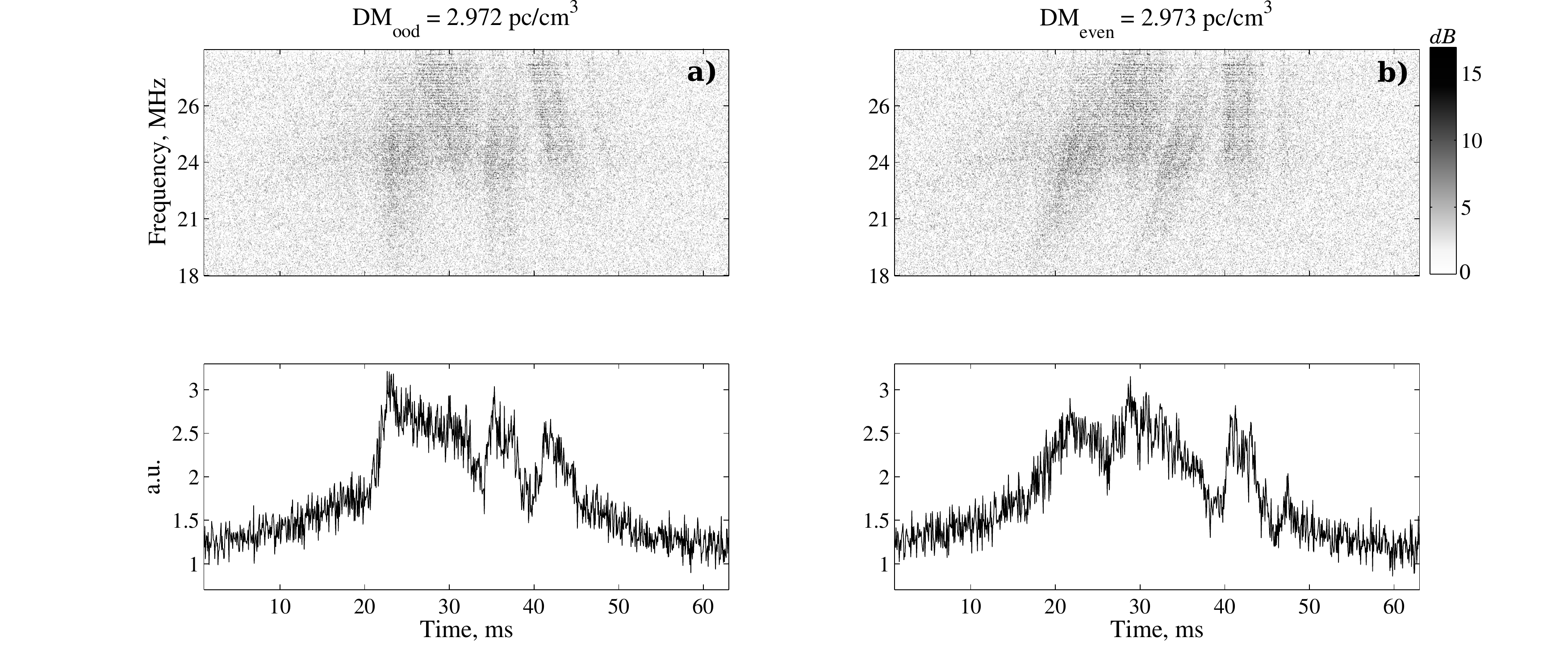} 
		\caption{The dynamic spectra and temporal profiles of the AIP of PSR J0953+0755 for a) $\mbox{DM}_{odd} = 2.9720 ~\mbox{pc}~\mbox{cm}^{-3}$; b) $\mbox{DM}_{even} = 2.9730 ~pc ~cm^{-3}$. Time resolution is $\Delta \tau_{DS} = 62 \mu \mbox{s}$, frequency range 18 -- 28 MHz.}
	\end{figure*}
	
	We obtained the four-component structure of the AIP of the PSR J0953+0755 (Fig. 3) and the visible DM values of these four components. We define “the visible dispersion measure” as classical DM in cold weakly anisotropic plasma in the ISM. This definition is convenient for describing the transformation of the AIP dynamic spectrum (see Fig. 3), but it is not very correct because the generated radio waves pass through pulsar magnetosphere, where the ultra relativistic electron-positron plasma with high magnetic strength exist. Alternative parameter to the “visible dispersion measure” could be a matrix of mutual time delays between different spectral components, but from our point of view it is inconvenient, because it would require a lot of additional explanations. In case of the pulsar magnetosphere one must use the permittivity tensor to solve the dispersion equation (see for example \cite{1979AuJPh..32...61M, volokitin1985waves, 1986FizPl..12.1233L, 1986ApJ...302..120A,  2006MNRAS.366.1539P, 2013PhyU...56..164B}). Since the correct interpretation of the observed effect using appropriate models is complicated, we will discuss it in the next paper with all necessary calculation. Here we will perform the qualitative analysis. 
	
	As we can see from the Fig. 1, Fig. 3, the detected signal has strong elliptical polarization. It means that this radiation was formed in some extended region of the pulsar magnetosphere where it obtains its polarization parameters. While propagating from the emission region through the pulsar magnetosphere the radiation undergoes linear polarization transformation (scattered on secondary plasma in strong dipole magnetic field). The amplitude of the normal modes of radiation changes, but the total energy of two modes stays constant at every spectrum frequency. From some height, which is higher than the generation region, the process of linear transformation of normal modes stops and the relative polarization parameters (total, linear and circular polarization degree) become fixed. This height is called the polarization-limiting radius \citep{1970pewp.book.....G, Zheleznyakov1997, 2013PhyU...56..164B}. Petrova estimated the polarization-limiting radius analytically for two extreme conditions \citep{2003A&A...408.1057P}. In this regard, the accurate analysis of our data will help us to determine some limits on magnetosphere parameters, or will disprove the theoretical conclusions.
	
	We use average values of the odd and even components that can be computed as visible $\mbox{DM}_{odd}= 2.9720 \pm 1.2 \cdot 10^{-4} ~\mbox{pc}~\mbox{cm}^{-3}$ and visible $\mbox{DM}_{even}= 2.9730 \pm 1.2 \cdot 10^{-4} ~\mbox{pc}~\mbox{cm}^{-3}$. The odd components are separated from the even components only by 6.25 ms. Both these facts mean that the detected difference in the $\mbox{DM}$ could not be caused by ISM, interplanetary medium (IPM) or the Earth ionosphere  perturbations. The only possible explanation of such short fluctuations is quickly varying plasma inside the pulsar magnetosphere or a nearby pulsar  wind.  
	
	The difference between two detected  $\mbox{DM} $ values is $\Delta DM= | DM_{odd}- DM_{even} | = 1 \cdot 10^{-3} \pm 1.2 \cdot 10^{-4} ~\mbox{pc}~\mbox{cm}^{-3}$. From the study of the Saturn lightning \citep{2013Icar..224...14K, 2013OAP..Mylostna, 2014RPRA..Mylostna} the average fluctuation of DM corresponding to the IPM is $\delta DM_{IPM} \sim 4 \cdot 10^{-5} \pm 1 \cdot 10^{-5} ~\mbox{pc}~\mbox{cm}^{-3}$. A similar to DM parameter of the total electron concentration (TEC, for the Earth ionosphere) is $\delta DM_{Ion} \sim 1 \cdot 10^{-6}~\mbox{pc}~\mbox{cm}^{-3}$ \citep{1981A&A....97..366A, 2007RPRA..Aframovich, 2007DokES.417.1444A, 2007RPRA..Lisachenki, 2008DokPh..53..211A, 2010RPRA..Zanimovskiy}. It means that the detected DM difference inside the individual pulsar pulse with an accuracy of $\sim 10^{-4}~\mbox{pc}~\mbox{cm}^{-3}$ is significant and we claim that we are able to resolve the pulsar magnetosphere in depth.
	
	To find the presence of temporally correlated fine structure in the AIP we analysed the component pairs separately as it was described in the previous section. The ACFs of intensity of the differential signal of both pairs are shown below. The breakpoint in the ACFs indicates the presence of fine structure of intensity with characteristic time-scales $\tau_{FS}$ of approximately 1 ms (see Fig. 4). This value fits well with characteristic scattering time of the pulsar J0953+0755 obtained in decametre wave range previously \citep{2012ARep...56..417U}. Space scale ($L_{cor}$) of the correlation that corresponds to the time interval $\tau_{FS}$ is $ L_{cor} \leq 300$ km. This interval characterizes the spatial scale of projection of emission region on the surface of polarization-limiting radii in the pulsar magnetosphere. The energy contribution of the short-scaled fine structure into the long-scaled components are 80 $\%$ for the odd components and 60 $\%$ for the even components.  
	
	\begin{figure}
		\includegraphics[width=0.5\textwidth]{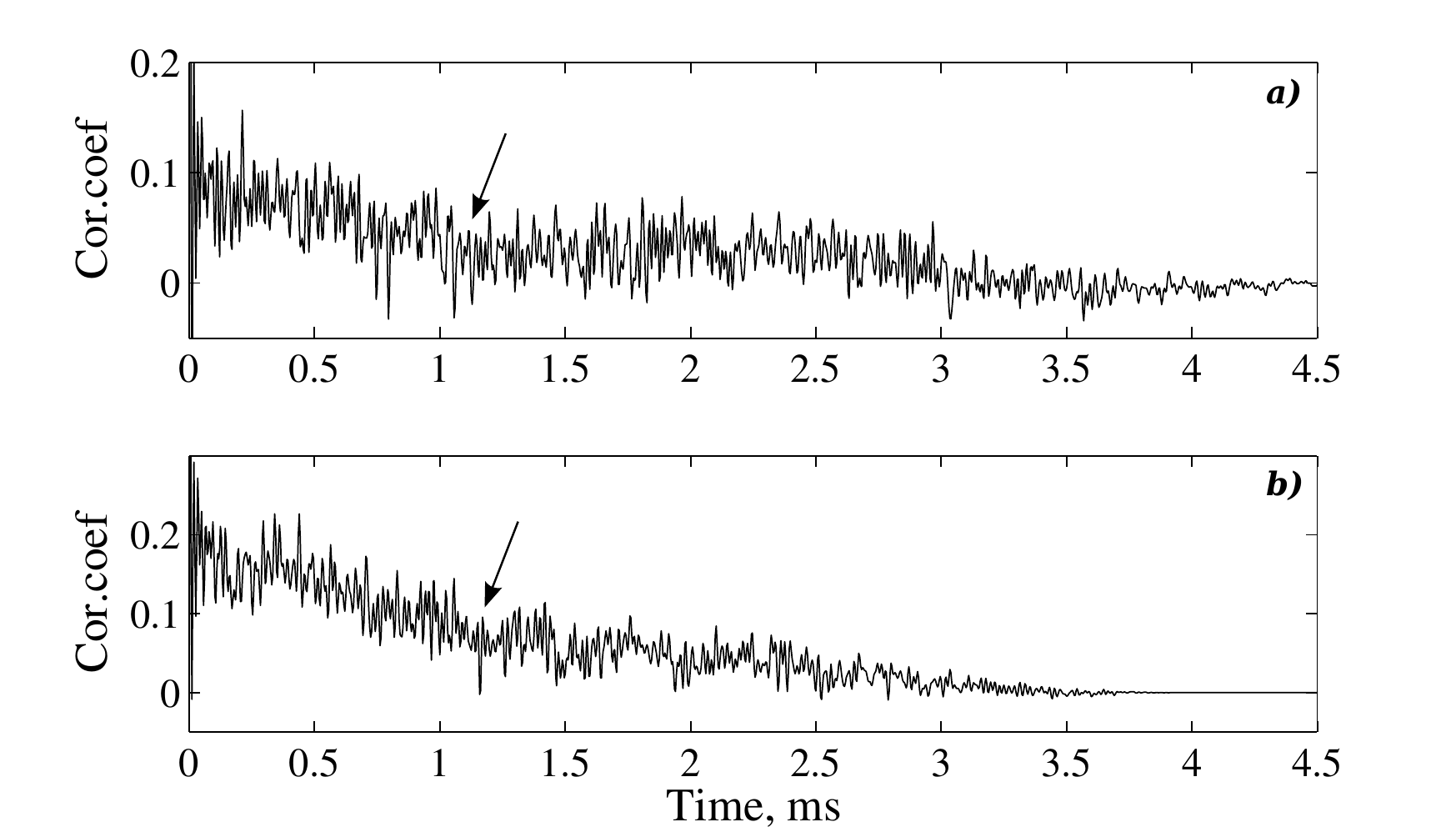} 
		\caption{a) The ACF of the odd components pair, $\mbox{DM}_{odd}= 2.9720 ~\mbox{pc}~\mbox{cm}^{-3}$; b) The ACF of the even components pair, $\mbox{DM}_{even}= 2.9730 ~\mbox{pc}~\mbox{cm}^{-3}$. The characteristic time-scale of the fine structure for both ACFs is close to 1 ms.}
	\end{figure}
	
	The resulting fine structure time-scale gives us another evidence in favour of the fact that the fine structure as well as difference corresponds to the effects of propagation through the pulsar magnetosphere. The characteristic time-scale of 1 ms corresponds to the effective spectral width of approximately 1 kHz. At the same time the characteristic bandwidth of IPM spectra fluctuations of scintillation is 1 Hz and scintillation on the ionospheric electron irregularities is 0.1 Hz \citep{2009Ap&SS.319..131K}. Considering that typical speed of electron concentration spatial irregularities movement is approximately 200 m $\mbox{s}^{-1}$ in the Earth ionosphere and $400 \cdot 10^3 $ m $\mbox{s}^{-1}$ in the IPM, the spatial scales of these irregularities are $\sim$ 0.2 m and $\sim$ 400 m respectively. The first value is much lower than the radiation wavelength, at the same time the scale of 400 m is much less than the dissipative scale of inhomogeneities in IPM \citep{2012RPRA....3..113O, 2012JASTP..86...34O}. 
	
	Also, we used ncross-correlation analysis to study correlative processes in separate pulse components and reduce noise influence. We obtained two properly dedispersed pulse component pairs as it was described earlier for ACFs constructions. The ncross-correlation function (CCF) of two pairs with different $\mbox{DM}s $ is shown in the Fig. 5. Here the CCF shows at least two scales of fine structure. The short-scale is 1 ms. It fits with the value found in the ACFs (Fig. 4). The long scale is 4 ms. Long-scale structure includes two or three short components. CCF shows that the odd and even pairs are separated from each other at $\Delta \tau_{\mathrm {CCF}} \approx 6.25$ ms.   
	
	\begin{figure}
		\includegraphics[width=0.5\textwidth]{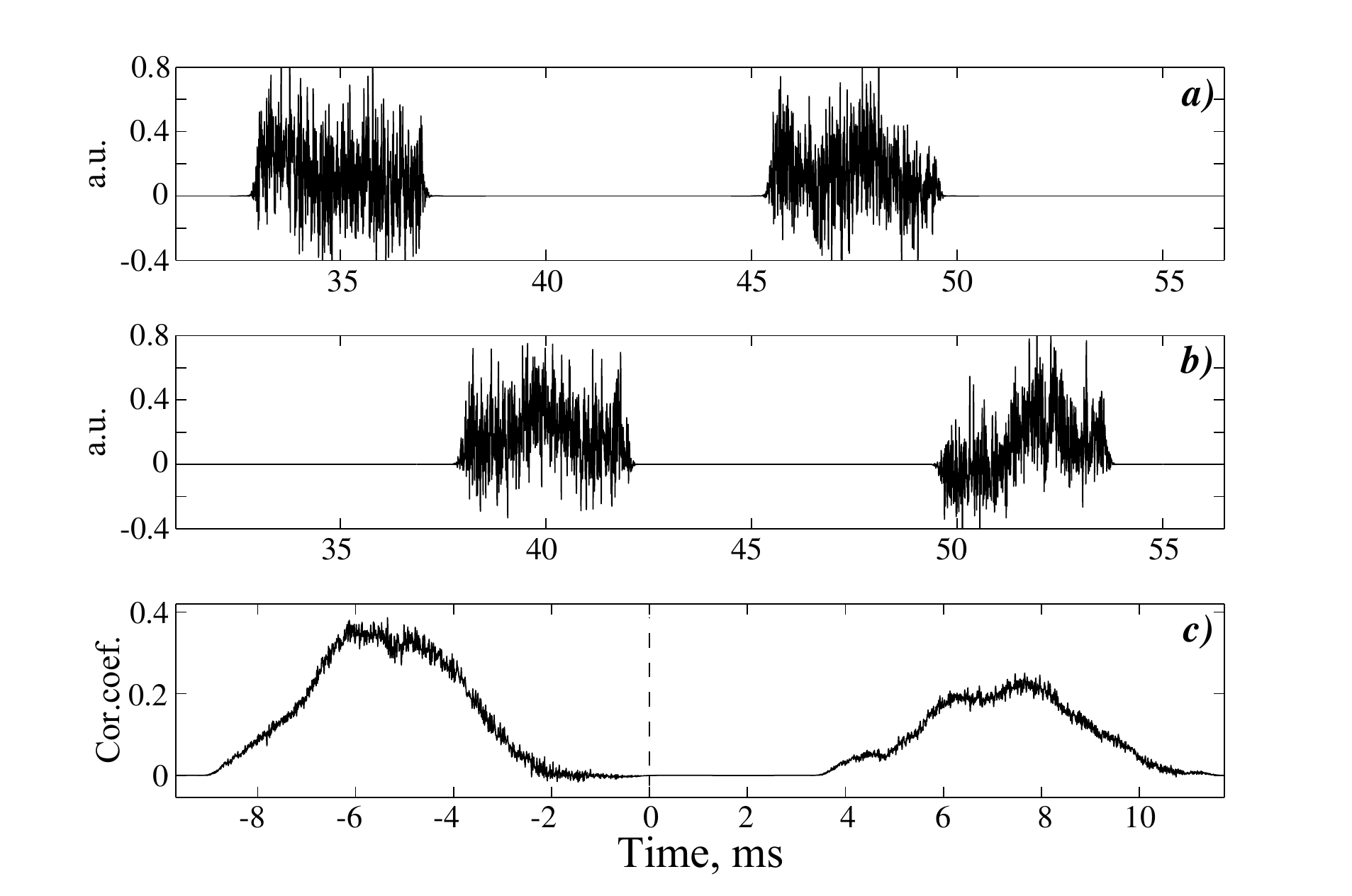} 
		\caption{a) The odd components pair, $\mbox{DM}_{odd}= 2.972 ~\mbox{pc}~\mbox{cm}^{-3}$; b) The even components pair, $\mbox{DM}_{even}= 2.973 ~\mbox{pc}~\mbox{cm}^{-3}$. c) The CCF of the two pairs shows two scales of fine structure: long-scale is 4 ms and short scale is 1 ms.}
	\end{figure}
	
	The presence of the radiation fine structure at decametre wavelengths allows us to estimate the Lorentz factor $\gamma$ of the ultra relativistic electrons and positrons emitting at these wavelengths.  The aberration compression of the radiation pattern of charge beam is proportional to the characteristic time of the ACF feature. On the other hand, the beam angle of the radiation pattern is inversely proportional to the Lorentz factor of emitting charges. Therefore one can write the equation: 
	\begin{eqnarray}   
	&&\gamma = {{1} \over {\sqrt{1 - \left({\dfrac{\nu}{c}}\right)^2}}} \approx {{P_{0}} \over {2 \pi \tau_{FS}} } \approx  40.27 ~,
	\end{eqnarray}
	where $\nu$ is the average charge speed,
	$P_0 \approx 0.253 ~\mbox{s}$ is the rotation period of PRS J0953+0755 taken from the ATNF data base \citep{2005AJ....129.1993M}.
	
	Let us give the equation for limit-polarization radius $R_{PL}$ from \citep{2006MNRAS.366.1539P} to clarify our explanation: 
	\begin{eqnarray}	
	&&\dfrac{\omega}{c}  \cdot (n_{O} - n_{X}) \Delta R_{PL} \approx 1; \nonumber \\ 
	&& R_{PL} 	\gg R_{gen} \Rightarrow  \Delta R_{PL} \approx R_{PL} ~,
	\end{eqnarray}
	where $\omega = 2 \pi f$ is cyclic frequency, $c$ is the speed of light, $(n_{O} - n_{X})$ is the difference between refraction coefficients of ordinary (O) and extraordinary (X) normal wave modes, $\Delta R_{PL}$ is the distance between generation zone $R_{gen}$ and polarization-limiting radius $R_{PL}$. 
	
	As far as all estimations include local plasma frequency, gyrotropic frequency, distance to the light cylinder and distance from a pulsar to ISM, the equations that describe these parameters are as follows:
	\begin{eqnarray}   
	&& \rho_{GJ} (r, \alpha, \varphi ) = - \frac{\mathbf{\Omega} \mathbf{B} (r, \alpha, \varphi )}{2 \pi c} \gamma_{\perp }(r, \alpha) = \nonumber \\ 
	&&~~~~~ - \frac{\Omega B (r, \alpha, \varphi ) \cos \alpha}{2 \pi c}  \frac{1}{\sqrt{1 - (\Omega r/ c)^2 \sin^2 \alpha}} ~,
	\end{eqnarray}
	where $\rho_{GJ} (r, \alpha, \varphi )$ is the Goldreich-Julian space-charge density \cite{1969ApJ...157..869G}, $ (r, \alpha, \varphi )$ are the polar coordinates in the local region on the pulsar magnetosphere ( $r$ is the radius, $\alpha$ is the polar angle, $\varphi$ is the azimuthal angle), $\mathbf{\Omega}$ is the rotation cyclic frequency vector, $\mathbf{B}(r, \alpha, \varphi )$ is the magnetic induction vector, $\gamma_{\perp }(r, \alpha)$ is the tangential Lorentz factor. 
	
	Local plasmas frequency can be estimated from the next equations:
	\begin{eqnarray}    
	&&\omega_{e^{-}, e^{+}}(r, \alpha, \varphi) = \sqrt{\frac{4 \pi e^2 N_{e^{-}, e^{+} }(r, 	\alpha, \varphi) }{m_e}} ~;\nonumber \\
	&&\omega_{SP}^2(r, \alpha, \varphi) = \sum_{q }\omega_{q}^2(r, \alpha, \varphi),
	\end{eqnarray} 
	where $\omega_{e^{-}, e^{+}}(r, \alpha, \varphi)$ are the  electron/positron local cyclic frequencies of the secondary pulsar plasma,
	$e$ is the electron charge,
	$N_{e^{-}, e^{+} }(r, \alpha, \varphi) $ are the electron/positron local concentrations, $m_e$ is the electron/positron rest mass,
	$\omega_{SP}(r, \alpha, \varphi)$ is the total secondary plasma cyclic frequency, 
	the index $q$ characterizes the particle charge ($e^-$ or $e^+ $).
	
	From equation (3) follows:
	\begin{eqnarray}    
	&& N_{SP}(r, \alpha, \varphi) = \sum_{q} N_{q}(r, \alpha, \varphi) \approx \left | \frac{K \cdot \rho_{GJ} (r, \alpha, \varphi)} {e} \right |,
	\end{eqnarray}
	where $N_{SP}(r, \alpha, \varphi)$ is the total secondary plasma charge concentration,
	$100 \leqslant K \leqslant 1000$ is the multiplication factor of the secondary plasma. 
	
	Gyrotropic frequencies are the same for electrons and positrons if Lorentz factors of electrons and positrons  are the same: 
	\begin{eqnarray}   
	&&\omega_{h}(r, \alpha, \varphi) = \frac {\left | {e \mathbf{B} (r, \alpha, \varphi)} \right |}{m_e c \gamma (r, \alpha, \varphi)},
	\end{eqnarray} 
	where $\omega_{h}(r, \alpha, \varphi)$ is the electron/positron local gyrotropic  frequency, 
	$\gamma (r, \alpha, \varphi)$ is the total Lorentz factor (see equation 1).
	
	We suggest that the dipole component of the magnetic field as well as the secondary electron--positron plasma concentration depends inversely as the cube of the distance from the pulsar surface.
	\begin{equation} 
	\left\{\begin{array}{l}		
	B(r, \alpha, \varphi) \approx B(R_{sur}, \alpha, \varphi) \cdot ({R_{sur}}/{r})^3\\
	N_{SP}(r, \alpha, \varphi) \approx N_{SP}(R_{sur}, \alpha, \varphi) \cdot({R_{sur}}/{r})^3\\
	r \leq R_{LC}/\sin \alpha; \ R_{LC} = c / \Omega 
	\end{array}\right.,	
	\end{equation}
	where $R_{LC} \approx 12 079$ km is the light cylinder radius, $R_{sur} = 10$ km is the neutron star radius.  
	
	The magnetic field vector on the neutron star surface is usually estimated by equating the magnetic-dipole losses to kinematic losses of rotating pulsar (we are talking about usual radio pulsars). We took the magnetic induction $B (R_{sur}, \alpha, \varphi) = 2.44 \cdot 10^{11}$ G from the ATNF pulsar data
	base \citep{2005AJ....129.1993M, atnfCite2014}.  The angle between the rotation axes and the magnetic induction vector is $\beta = \angle (\mathbf \Omega, \mathbf B) \approx 15^{\circ}$  taken as average from \citep{2011ARep...55...19M}. To simplify we can define the distance from the pulsar centre to the light cylinder surface for the chosen polar angle $\alpha$: ${R}'_{LC}(\alpha) = R_{LC} / \sin \alpha$ .
	
	At distance $r > {R}'_{LC} (\alpha) $ we assumed (see equation 7) the magnetic induction to change inversely on the distance from the neutron star centre and the secondary plasma concentration to change inversely as the square of the distance:
	\begin{equation}   
	\left\{\begin{array}{l}		
	B(r, \alpha, \varphi) \approx B(R'_{LC}(\alpha), \alpha, \varphi) \cdot {R'_{LC}(\alpha)} / {r} \\
	N_{SP}(r, \alpha, \varphi) \approx N_{SP} (R'_{LC}(\alpha), \alpha, \varphi) \cdot ({R'_{LC}(\alpha)} / {r})^2 \\
	r > R'_{LC}(\alpha)
	\end{array}\right..	
	\end{equation}
	Fig. 6 presents the dependences of the $N_{SP}(r, \beta, \varphi)$, $\omega_{SP} (r, \beta, \varphi)$ and $\omega_h (r, \beta, \varphi)$ versus distance from the pulsar. 
	
	\begin{figure}          
		\begin{center}
			\includegraphics[width=0.4\textwidth]{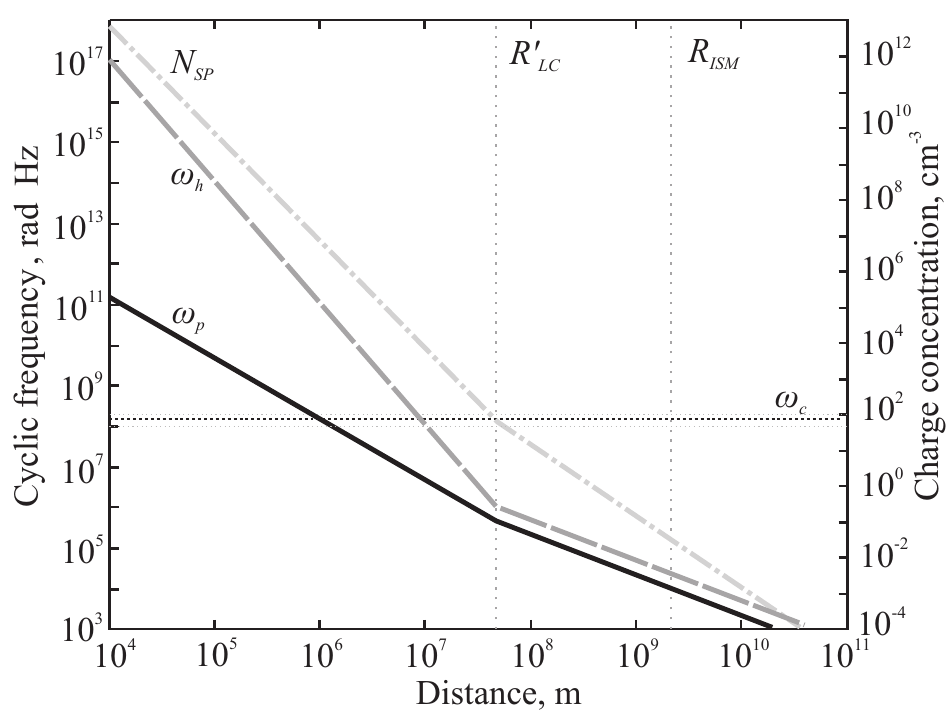} 
			\caption{The charge particle concentration $N_{SP}(r, \beta, \varphi)$ of secondary electron-positron plasma (see equations (5,7,8)), gyrotropic frequency $\omega_h$ (see equations 6,7,8) and plasma frequency $\omega_{SP}$ (see equations 4,7,8) of the PSR J0953+0755 versus distance from the pulsar. Vertical lines show the distance between the pulsar center and the light cylinder surface $R'_{LC}$ and the suppositive distance to the ISM $R_{ISM}$. For value  $\beta = 15^{\circ} $ is the angle between $\mathbf{\Omega}$ and $\mathbf{B}$ vectors.}
		\end{center}
	\end{figure}
	
	The distance from the pulsar to the ISM $R_{ISM}(\beta)$ could be computed from the condition of equality of the charge concentration in the pulsar wind to average electron concentration in the ISM $N_{SP}(R_{ISM}(\beta), \beta, \varphi) = ~<N_{ISM}> \ \approx \ 0.03 ~\mbox{cm}^{-3}$.
	
	The most interesting is to estimate the polarization-limiting radius of the detected fine structure of the PSR J0953+0755 radiation. The angle of the radiation beam related to the fine structure is $\delta = 2 \pi \tau_{FS} / P_0 = 1 / \gamma$. From the paper \cite{2006MNRAS.366.1539P} the relative polarization-limiting radius can be estimated as the following:
	\begin{eqnarray}  
	\frac {R_{PL}(K, \theta)} {R_{LC}} = \frac{0.018 \cdot \! P_{0}^{\!-\tfrac{3}{2}}} {\theta} \left ( \frac {\gamma}{10^{1.5}} \right )^{\!\!-\tfrac{3}{2}} 
	\!\! \sqrt{\frac{K}{100} \frac{B_{SUR}}{10^{12}} \frac {10^9}{f_c}}~ ,
	\end{eqnarray}
	where $\gamma = P_{0}/ (2 \pi \tau_{FS})$ is the Lorentz factor of secondary plasma related to the fine structure $\tau_{FS} = 1$ ms, $P_0$ is the pulsar rotation period, $K$ is the multiplication factor, $f_c$ is the central frequency, $\theta$ is the angle between the wave vector $\mathbf{k(\omega)}$ and the ambient magnetic field.	
	
	The solutions of the equation (9) with parameters corresponding to the PSR  J0953+0755 are given in the Fig. 7. These solutions are given in the parametric form under conditions: $K \in [100; 1000]; \ \theta \in [10^{\circ}; 90^{\circ}]; R_{gen}(K) \leq R_{PL}( K, \theta ) \leq R'_{LC}; \ f_c = 24$ MHz.  
	
	The range of polarization-limiting radii  (equation 9) depends on pulsar magnetosphere geometry, magnetosphere parameters and on the frequency of observation. It means that the observational results have great potential for probing the upper layers of the pulsar magnetosphere in depth.  One should limit the polarization-limiting radius surface, but it is not necessary for qualitative understanding. If we replace $R_{PL}(K, \theta)$ in equation (9) by $\Delta R_{PL}(K, \theta)$, then the limit $R_{PL}(K, \theta) \gg R_{gen} (K)$ from equation (2), we can alleviate condition and write $R_{PL}(K, \theta) \geq R_{gen}(K)$.  
	
	\begin{figure*}
		\begin{center}
			\includegraphics[width=1\textwidth]{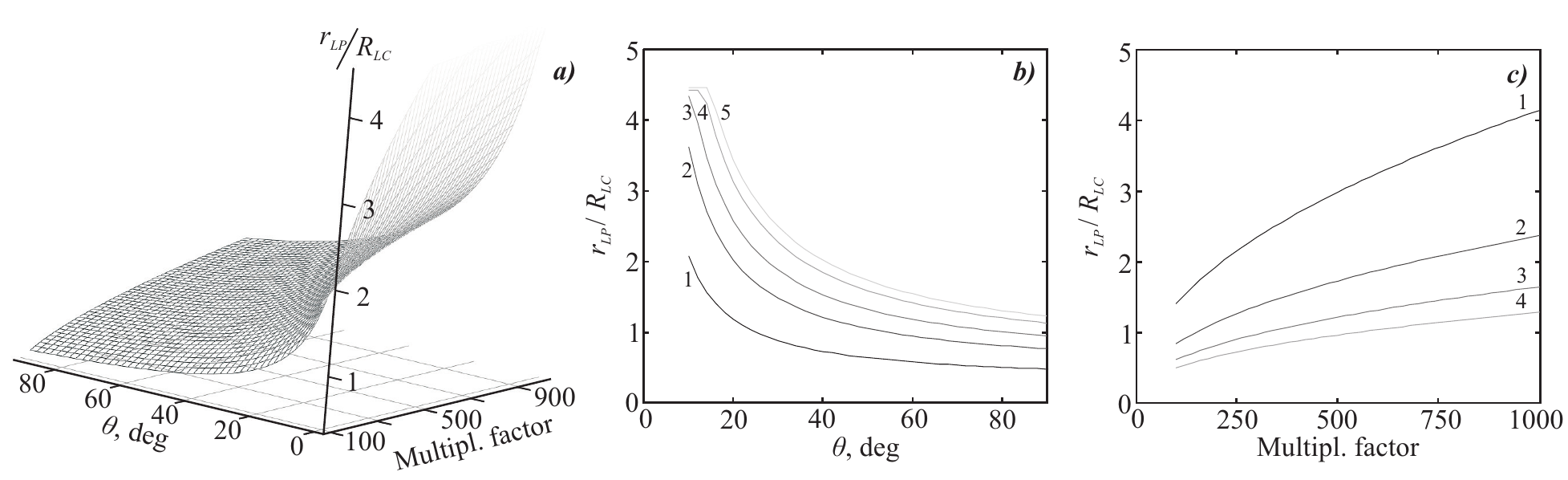} 
			\caption{a) The polarization-limiting radii surface dependence on the angle between the wave vector and the ambient magnetic field $\theta \in [10^{\circ}; 90^{\circ}]$ and the multiplication factor $K \in [100; 1000]$ at central frequency $f_c = 24$ MHz for the short scale of the fine structure of  PSR J0953+0755 (see equation (9)). The area of possible solutions is limited to the top by $R_{PL}(\omega_c, K, \theta) \leq R'_{LC}(\beta)$ (where $\beta = 15^{\circ}$); b) the cross-section of the surface in a plane of $K$, where the numbers 1, 2, 3, 4, 5 correspond to the multiplication factors $K = 100, 320, 540, 820, 1000$; c) the cross-section of the surface in a plane of $\theta$, where the numbers 1, 2, 3, 4 correspond to the angles $\theta = 16^{\circ}, 32^{\circ}, 54^{\circ}, 82^{\circ}$.}
		\end{center}
	\end{figure*}
	
	We believe that our reasoning allow evaluation of qualitative parameters of pulsar radiation in polarization-limiting region. The closer the propagation conditions to the transverse ($\theta = \pi / 2$) and the less the multiplication factor is, the closer the critical radius is pressed to the pulsar surface. Conversely, the bigger the multiplication factor and the less the angle between the wave vector and the magnetic field are, the higher from the pulsar surface the polarization-limiting radius is. 
	
	The last equation (equation 9) allows us to make rough estimates of the polarization-limiting radii. But it shows that the altitudes at which the emission finish the polarization evolution, are located inside the light cylinder even in the decametre range. This gives reason to continue attempts to resolve the pulsar magnetosphere in depth. 
	
	The possible explanations of the detected difference in propagation time (visible dispersion measures) of the odd and even components of the AIP of PSR J0953+0755 are as follows. 
	\begin{enumerate}
		\item The components were radiated from different regions of the pulsar magnetosphere, for example from polar and outer gaps. We can roughly estimate distance between emission regions as $\Delta r = c \cdot \Delta \tau_{\mathrm {CCF}} = 1875$ km, where $\tau_{\mathrm {CCF}} = 6.25$ ms is the time interval between two pairs of components in the dynamic spectrum (see Fig. 1, Fig. 3).   
		
		\item The signals split into ordinary and extraordinary wave modes during propagating from the emission region to the polarization-limiting radius due to the birefringence. Wherein, two component pairs connected with ordinary and extraordinary waves are experiencing linear transformation of polarization parameters. The birefringence was detected for several pulsars in \citep{2015A&A...576A..62N}. In decametre wave range this effect must be the most contrasted.       
	\end{enumerate}
	
	In our case both pairs of the pulse components are elliptically polarized. The fact that these components are the normal modes of electromagnetic waves travelling through pulsar magnetosphere with strong magnetic field \citep{2013BaltA..22...53U} can be the cause of the elliptic polarization. The magnetic field of the PSR J0953+0755 in the visible emission region is strong enough under our consideration. Moreover, while the fine structure has strong linear polarization, the linear polarization of radiation of the main pulse (about 50 ms long) is much weaker. Probably, the polarization conditions of fine structure and other parts of 'ordinary' pulse are different and they are radiated from different regions in a pulsar magnetosphere. This fact can confirm the hypothesis of the birefringence detection in decametre wave range. Formally, time of registration (visible dispersion measure) of the ordinary and extraordinary waves must be different not only because of birefringence, but also because of the different polarization-limiting radii for each of the components. To get additional information about polarization parameters, one should observe the pulsars at a radio telescope that consists of orthogonal dipoles (such as URAN-2 or LOFAR) or that has right-hand and left-hand circular polarization modes (such as Nan\c{c}ay decametre Array).  
	
	We believe that the detected fine structure of the PSR J0953+0755 decametre radiation is explained by propagation effects in the pulsar magnetosphere, but not as a result of neutron star seismic modulation.

	\section{Conclusions}
	Four components inside the AIP of the pulsar J0953+0755 were detected in the decametre wave range. The odd and even pairs of the components have different visible dispersion measures. The obtained difference is reliable and relates to the plasma perturbation in the pulsar magnetosphere. 
	
	We detected several scales of fine structure of the AIP at 20 -- 28 MHz. The characteristic time-scale of the shortest structure is 1 ms for both component pairs. The long-scale is approximately 4 ms. We claim that the fine structure of both time scales and ordinary pulsar radiation of the main pulse are probably registrated from different height at decametre wavelengths. The short structure is likely to be emitted from several narrow magnetic tubes in the pulsar magnetosphere.   
	
	We estimated the Lorenz factor of the ultra relativistic electrons and positrons in the polarization-limiting region. 
	
	The detection of differential in visible DM at millisecond time-scale is a good prerequisites to resolve the pulsar magnetosphere in depth and along azimuthal angle. 
	
	The decametre wavelengths are promising to study fine effects of the propagation medium because the relative frequency band $\Delta f / f_c$ is the biggest at low frequencies and all effects are most contrast in the stable propagation conditions. Thus the area of the studies is newsworthy and the biggest radio telescopes in the decametre and meter range should be involved into these studies.
	
	\section{Acknowledgments}
	
	These researches are supported by grant PICS 1.33.11. The authors would like to acknowledge the valuable comments and suggestions of G. I. Melikidze, which have improved the quality of this paper. The authors would like to thank their colleagues S. A. Petrova for many useful comments and discussions on the topic and I.Y. Vasylieva for a critical reading of the original version of the paper.

	\bibliography{FSDM_a} 

\begin{thebibliography}{71}
\expandafter\ifx\csname natexlab\endcsname\relax\def\natexlab#1{#1}\fi

\bibitem[{{Afraimovich}(1981)}]{1981A&A....97..366A}
{Afraimovich} E.~L., 1981, A\&A, 97, 366

\bibitem[{{Afraimovich}(2007)}]{2007DokES.417.1444A}
{Afraimovich} E.~L., 2007, Doklady Earth Sciences, 417, 1444

\bibitem[{{Afraimovich}, {Smol'Kov} \& {Yasyukevich}(2008){Afraimovich},
  {Smol'Kov}, \& {Yasyukevich}}]{2008DokPh..53..211A}
{Afraimovich} {\'E}.~L., {Smol'Kov} G.~Y., {Yasyukevich} Y.~V., 2008, Physics -
  Doklady, 53, 211

\bibitem[{{Afraimovich} \& {Yasyukevich}(2007)}]{2007RPRA..Aframovich}
{Afraimovich} E.~L., {Yasyukevich} Y.~V., 2007, Radio Physics and Radio
  Astronomy, 12, 357, in russian

\bibitem[{{Arons} \& {Barnard}(1986)}]{1986ApJ...302..120A}
{Arons} J., {Barnard} J.~J., 1986, ApJ, 302, 120

\bibitem[{{Beskin}, {Istomin} \& {Philippov}(2013){Beskin}, {Istomin}, \&
  {Philippov}}]{2013PhyU...56..164B}
{Beskin} V.~S., {Istomin} Y.~N., {Philippov} A.~A., 2013, Physics Uspekhi, 56,
  164

\bibitem[{{Boriakoff}(1976)}]{1976ApJ...208L..43B}
{Boriakoff} V., 1976, ApJ, 208, L43

\bibitem[{{Boriakoff} \& {Ferguson}(1981)}]{1981IAUS...95..191B}
{Boriakoff} V., {Ferguson} D.~C., 1981, in IAU Symposium, Vol.~95, Pulsars: 13
  Years of Research on Neutron Stars, {Sieber} W., {Wielebinski} R., eds., pp.
  191--196

\bibitem[{{Boriakoff}, {Ferguson} \& {Slater}(1981){Boriakoff}, {Ferguson}, \&
  {Slater}}]{1981IAUS...95..199B}
{Boriakoff} V., {Ferguson} D.~C., {Slater} G., 1981, in IAU Symposium, Vol.~95,
  Pulsars: 13 Years of Research on Neutron Stars, {Sieber} W., {Wielebinski}
  R., eds., pp. 199--204

\bibitem[{{Drake} \& {Craft}(1968)}]{1968Natur.220..231D}
{Drake} F.~D., {Craft} H.~D., 1968, Nature, 220, 231

\bibitem[{{Ershov} \& {Kuzmin}(2003)}]{2003AstL...29...91E}
{Ershov} A.~A., {Kuzmin} A.~D., 2003, Astronomy Letters, 29, 91

\bibitem[{{Gil} \& {Melikidze}(2004)}]{2004IAUS..218..321G}
{Gil} J., {Melikidze} G.~I., 2004, in IAU Symposium, Vol. 218, Young Neutron
  Stars and Their Environments, {Camilo} F., {Gaensler} B.~M., eds., CA:
  Astronomical Society of the Pacific, pp. 321--324

\bibitem[{{Gil}, {Melikidze} \& {Geppert}(2003){Gil}, {Melikidze}, \&
  {Geppert}}]{2003A&A...407..315G}
{Gil} J., {Melikidze} G.~I., {Geppert} U., 2003, A\&A, 407, 315

\bibitem[{{Ginzburg}(1970)}]{1970pewp.book.....G}
{Ginzburg} V.~L., 1970, {The propagation of electromagnetic waves in plasmas},
  2nd edn. Pergamon, Oxford

\bibitem[{{Goldreich} \& {Julian}(1969)}]{1969ApJ...157..869G}
{Goldreich} P., {Julian} W.~H., 1969, ApJ, 157, 869

\bibitem[{{Hankins}(1971)}]{1971ApJ...169..487H}
{Hankins} T.~H., 1971, ApJ, 169, 487

\bibitem[{{Hankins}(1972)}]{1972ApJ...177L..11H}
{Hankins} T.~H., 1972, ApJ, 177, L11

\bibitem[{{Hankins} \& {Eilek}(2007)}]{2007ApJ...670..693H}
{Hankins} T.~H., {Eilek} J.~A., 2007, ApJ, 670, 693

\bibitem[{{Hankins} {et~al}\mbox{.}(2003){Hankins}, {Kern}, {Weatherall}, \&
  {Eilek}}]{2003Natur.422..141H}
{Hankins} T.~H., {Kern} J.~S., {Weatherall} J.~C., {Eilek} J.~A., 2003, Nature,
  422, 141

\bibitem[{{Hankins} \& {Rickett}(1975)}]{1975MComP..14...55H}
{Hankins} T.~H., {Rickett} B.~J., 1975, Methods in Computational Physics, 14,
  55

\bibitem[{{Hobbs}, {Manchester} \& {Toomey}(2015){Hobbs}, {Manchester}, \&
  {Toomey}}]{atnfCite2014}
{Hobbs} G., {Manchester} R.~N., {Toomey} L., 2015, {ATNF Pulsar Catalog}.
  http://www.atnf.csiro.au/people/pulsar/psrcat/

\bibitem[{{Kalinichenko}(2009)}]{2009Ap&SS.319..131K}
{Kalinichenko} N.~N., 2009, Ap\&SS, 319, 131

\bibitem[{{Konovalenko} {et~al}\mbox{.}(2013){Konovalenko}, {Kalinichenko},
  {Rucker}, {Lecacheux}, {Fischer}, {Zarka}, {Zakharenko}, {Mylostna},
  {Grie{\ss}meier}, {Abranin}, {Falkovich}, {Sidorchuk}, {Kurth}, {Kaiser}, \&
  {Gurnett}}]{2013Icar..224...14K}
{Konovalenko} A.~A. {et~al.}, 2013, Icarus, 224, 14

\bibitem[{{Kontorovich} \& {Flanchik}(2007)}]{2007JETPL..85..267K}
{Kontorovich} V.~M., {Flanchik} A.~B., 2007, Soviet Journal of Experimental and
  Theoretical Physics Letters, 85, 267

\bibitem[{{Kontorovich} \& {Flanchik}(2013)}]{2013Ap&SS.345..169K}
{Kontorovich} V.~M., {Flanchik} A.~B., 2013, Ap\&SS, 345, 169

\bibitem[{{Lisachenko} {et~al}\mbox{.}(2007){Lisachenko}, {Zanimonskiy},
  {Yampolsky}, \& {Wielgosz}}]{2007RPRA..Lisachenki}
{Lisachenko} V.~N., {Zanimonskiy} Y.~M., {Yampolsky} Y.~M., {Wielgosz} P.,
  2007, Radio Physics and Radio Astronomy, 12, 20

\bibitem[{{Lominadze} {et~al}\mbox{.}(1986){Lominadze}, {Machabeli},
  {Melikidze}, \& {Pataraia}}]{1986FizPl..12.1233L}
{Lominadze} D.~G., {Machabeli} G.~Z., {Melikidze} G.~I., {Pataraia} A.~D.,
  1986, Fizika Plazmy, 12, 1233

\bibitem[{{Malofeev}, {Malov} \& {Shchegoleva}(1998){Malofeev}, {Malov}, \&
  {Shchegoleva}}]{1998ARep...42..241M}
{Malofeev} V.~M., {Malov} O.~I., {Shchegoleva} N.~B., 1998, Astronomy Reports,
  42, 241

\bibitem[{{Malov} \& {Nikitina}(2011)}]{2011ARep...55...19M}
{Malov} I.~F., {Nikitina} E.~B., 2011, Astronomy Reports, 55, 19

\bibitem[{{Manchester} {et~al}\mbox{.}(2005){Manchester}, {Hobbs}, {Teoh}, \&
  {Hobbs}}]{2005AJ....129.1993M}
{Manchester} R.~N., {Hobbs} G.~B., {Teoh} A., {Hobbs} M., 2005, AJ, 129, 1993

\bibitem[{{McLaughlin} {et~al}\mbox{.}(2006){McLaughlin}, {Lyne}, {Lorimer},
  {Kramer}, {Faulkner}, {Manchester}, {Cordes}, {Camilo}, {Possenti}, {Stairs},
  {Hobbs}, {D'Amico}, {Burgay}, \& {O'Brien}}]{2006Natur.439..817M}
{McLaughlin} M.~A. {et~al.}, 2006, Nature, 439, 817

\bibitem[{{Megn} {et~al}\mbox{.}(1978){Megn}, {Sodin}, {Sharykin}, {Bruk},
  {Melyanovsky}, {Inyutin}, \& {Goncharov}}]{1978MenSodin}
{Megn} A.~V., {Sodin} L.~G., {Sharykin} N.~K., {Bruk} Y.~M., {Melyanovsky}
  P.~A., {Inyutin} G.~A., {Goncharov} N.~U., 1978, in Antennas, Vol.~26,
  Antennas (Collection of papers), A. P.~A., ed., Scientific-Technical Society
  of Radio Engineering, Electronics and Communications. AS, Popova; Svyaz,
  Moscow, pp. 15--57

\bibitem[{{Melrose}(1979)}]{1979AuJPh..32...61M}
{Melrose} D.~B., 1979, Australian Journal of Physics, 32, 61

\bibitem[{{Mylostna} {et~al}\mbox{.}(2013){Mylostna}, {Zakharenko},
  {Konovalenko}, {Kolyadin}, {Zarka}, {Gremeier}, {Litvinenko}, {Sidorchuk},
  {Rucker}, {Fischer}, {Cecconi}, {Nikolaenko}, \& V.}]{2013OAP..Mylostna}
{Mylostna} K. {et~al.}, 2013, Odessa Astronomical Publications, 97, 251

\bibitem[{{Mylostna} {et~al}\mbox{.}(2014){Mylostna}, {Zakharenko},
  {Konovalenko}, {Fischer}, {Zarka}, \& {Sidorchuk}}]{2014RPRA..Mylostna}
{Mylostna} K., {Zakharenko} V., {Konovalenko}, A. V., {Fischer} G., {Zarka} P.,
  {Sidorchuk} M., 2014, Radio Physics and Radio Astronomy, 19, 10

\bibitem[{{Noutsos} {et~al}\mbox{.}(2015){Noutsos}, {Sobey}, {Kondratiev},
  {Weltevrede}, {Verbiest}, {Karastergiou}, {Kramer}, {Kuniyoshi}, {Alexov},
  {Breton}, {Bilous}, {Cooper}, {Falcke}, {Grie{\ss}meier}, {Hassall},
  {Hessels}, {Keane}, {Os{\l}owski}, {Pilia}, {Serylak}, {Stappers}, {ter
  Veen}, {van Leeuwen}, {Zagkouris}, {Anderson}, {B{\"a}hren}, {Bell},
  {Broderick}, {Carbone}, {Cendes}, {Coenen}, {Corbel}, {Eisl{\"o}ffel},
  {Fender}, {Garsden}, {Jonker}, {Law}, {Markoff}, {Masters}, {Miller-Jones},
  {Molenaar}, {Osten}, {Pietka}, {Rol}, {Rowlinson}, {Scheers}, {Spreeuw},
  {Staley}, {Stewart}, {Swinbank}, {Wijers}, {Wijnands}, {Wise}, {Zarka}, \&
  {van der Horst}}]{2015A&A...576A..62N}
{Noutsos} A. {et~al.}, 2015, A\&A, 576, A62

\bibitem[{{Novikov} {et~al}\mbox{.}(1984){Novikov}, {Popov}, {Soglasnov},
  {Bruk}, \& {Ustimenko}}]{1984AZh....61..343N}
{Novikov} A.~Y., {Popov} M.~V., {Soglasnov} V.~A., {Bruk} Y.~M., {Ustimenko}
  B.~Y., 1984, Azh, 61, 343

\bibitem[{{Olyak}(2011)}]{2012RPRA....3..113O}
{Olyak} M.~R., 2011, Radio Physics and Radio Astronomy, 16, 366

\bibitem[{{Olyak}(2012)}]{2012JASTP..86...34O}
{Olyak} M.~R., 2012, Journal of Atmospheric and Solar-Terrestrial Physics, 86,
  34

\bibitem[{{Petrova}(2001)}]{2001A&A...378..883P}
{Petrova} S.~A., 2001, A\&A, 378, 883

\bibitem[{{Petrova}(2003)}]{2003A&A...408.1057P}
{Petrova} S.~A., 2003, A\&A, 408, 1057

\bibitem[{{Petrova}(2004)}]{2004A&A...417L..29P}
{Petrova} S.~A., 2004, A\&A, 417, L29

\bibitem[{{Petrova}(2006{\natexlab{a}})}]{2006MNRAS.366.1539P}
{Petrova} S.~A., 2006{\natexlab{a}}, MNRAS, 366, 1539

\bibitem[{{Petrova}(2006{\natexlab{b}})}]{2006MNRAS.368.1764P}
{Petrova} S.~A., 2006{\natexlab{b}}, MNRAS, 368, 1764

\bibitem[{{Popov}, {Smirnova} \& {Soglasnov}(1987){Popov}, {Smirnova}, \&
  {Soglasnov}}]{1987SvA....31..529P}
{Popov} M.~V., {Smirnova} T.~V., {Soglasnov} V.~A., 1987, SvA, 31, 529

\bibitem[{{Rankin}(1986)}]{1986ApJ...301..901R}
{Rankin} J.~M., 1986, ApJ, 301, 901

\bibitem[{{Rankin}(2013)}]{2013mnras...Rankin}
{Rankin}, J.~M. W. G. E. . B. A.~M., 2013, MNRAS, 433, 445

\bibitem[{{Rickett}(1975)}]{1975ApJ...197..185R}
{Rickett} B.~J., 1975, ApJ, 197, 185

\bibitem[{{Ryabov} {et~al}\mbox{.}(2010){Ryabov}, {Vavriv}, {Zarka}, {Ryabov},
  {Kozhin}, {Vinogradov}, \& {Denis}}]{2010A&A...510A..16R}
{Ryabov} V.~B., {Vavriv} D.~M., {Zarka} P., {Ryabov} B.~P., {Kozhin} R.,
  {Vinogradov} V.~V., {Denis} L., 2010, A\&A, 510, A16

\bibitem[{{Sidorchuk} {et~al}\mbox{.}(2008){Sidorchuk}, {Ulyanov}, {Shepelev},
  {Brazhenko}, {Vashchishin}, \& {Frantzusenko}}]{Sidorchuk2008}
{Sidorchuk} M.~A., {Ulyanov} O.~M., {Shepelev}, V. A. amd~{Mukha} D.~V.,
  {Brazhenko} A.~I., {Vashchishin} R.~V., {Frantzusenko} A.~V., 2008, in
  Scientific Workshop – Astrophysics with E-LOFAR

\bibitem[{{Smirnova}, {Tul'bashev} \& {Boriakoff}(1994){Smirnova},
  {Tul'bashev}, \& {Boriakoff}}]{1994A&A...286..807S}
{Smirnova} T.~V., {Tul'bashev} S.~A., {Boriakoff} V., 1994, A\&A, 286, 807

\bibitem[{{Soglasnov}(2000)}]{2000ASPC..202..181S}
{Soglasnov} V.~A., 2000, in Astronomical Society of the Pacific Conference
  Series, Vol. 202, IAU Colloq. 177: Pulsar Astronomy - 2000 and Beyond,
  {Kramer} M., {Wex} N., {Wielebinski} R., eds., p. 181

\bibitem[{{Soglasnov}, {Popov} \& {Kuzmin}(1983){Soglasnov}, {Popov}, \&
  {Kuzmin}}]{1983SvA....27..169S}
{Soglasnov} V.~A., {Popov} M.~V., {Kuzmin} O.~A., 1983, SvA, 27, 169

\bibitem[{{Soglasnov} {et~al}\mbox{.}(2001){Soglasnov}, {Skulachev}, {D'Amico},
  {Montebugnoli}, {Semenkov}, {Maccaferri}, \& {Cattani}}]{2001ARep...45..294S}
{Soglasnov} V.~A., {Skulachev} A.~D., {D'Amico} N., {Montebugnoli} S.,
  {Semenkov} K.~V., {Maccaferri} A., {Cattani} A., 2001, Astronomy Reports, 45,
  294

\bibitem[{{Soglasnov} {et~al}\mbox{.}(1981){Soglasnov}, {Smirnova}, {Popov}, \&
  {Kuzmin}}]{1981SvA....25..442S}
{Soglasnov} V.~A., {Smirnova} T.~V., {Popov} M.~V., {Kuzmin} A.~D., 1981, SvA,
  25, 442

\bibitem[{{Tsai} {et~al}\mbox{.}(2015){Tsai}, {Simonetti}, {Akukwe}, {Bear},
  {Cutchin}, {Dowell}, {Gough}, {Kanner}, {Kassim}, {Schinzel}, {Shawhan},
  {Taylor}, {Yancey}, {Quezada}, \& {Kavic}}]{2015AJ....149...65T}
{Tsai} J.-W. {et~al.}, 2015, AJ, 149, 65

\bibitem[{{Ulyanov} {et~al}\mbox{.}(2006){Ulyanov}, { Zakharenko},
  {Konovalenko}, {Lecacheux}, {Rosolen}, \& {Rucker}}]{UlyanovZaharenko2006}
{Ulyanov} O.~M., { Zakharenko} V.~V., {Konovalenko} A.~A., {Lecacheux} A.,
  {Rosolen} K., {Rucker} H.~O., 2006, Radio Physics and Radio Astronomy, 11,
  113

\bibitem[{{Ulyanov} {et~al}\mbox{.}(2007){Ulyanov}, {Deshpande}, {Zakharenko},
  {Asgekar}, \& {Shankar}}]{UlyanovDeshpande2007}
{Ulyanov} O.~M., {Deshpande} A., {Zakharenko} V.~V., {Asgekar} A., {Shankar}
  U., 2007, Radio Physics and Radio Astronomy, 12, 5

\bibitem[{{Ulyanov}, {Seredkina} \& {Shevtsova}(2013){Ulyanov}, {Seredkina}, \&
  {Shevtsova}}]{2013IAUS..Ulyanov}
{Ulyanov} O.~M., {Seredkina} A.~A., {Shevtsova} A.~I., 2013, in IAU Symposium,
  Vol. 291, IAU Symposium, {van Leeuwen} J., ed., pp. 530--532

\bibitem[{{Ulyanov} {et~al}\mbox{.}(2013){Ulyanov}, {Shevtsova}, {Mukha}, \&
  {Seredkina}}]{2013BaltA..22...53U}
{Ulyanov} O.~M., {Shevtsova} A.~I., {Mukha} D.~V., {Seredkina} A.~A., 2013,
  Baltic Astronomy, 22, 53

\bibitem[{{Ulyanov}, {Shevtsova} \& {Skoryk}(2013){Ulyanov}, {Shevtsova}, \&
  {Skoryk}}]{UlyanovShevtsova2013}
{Ulyanov} O.~M., {Shevtsova} A.~I., {Skoryk} A.~O., 2013, Bulletin of the
  Crimean Astrophysical Observatory, 109, 159

\bibitem[{{Ulyanov}, {Shevtsova} \& {Skoryk}(2014){Ulyanov}, {Shevtsova}, \&
  {Skoryk}}]{UlyanovShevtsova2014}
{Ulyanov} O.~M., {Shevtsova} A.~I., {Skoryk} A.~O., 2014, Radio Physics and
  Radio Astronomy, 19, 101

\bibitem[{{Ul'yanov} \& {Zakharenko}(2012)}]{2012ARep...56..417U}
{Ul'yanov} O.~M., {Zakharenko} V.~V., 2012, Astronomy Reports, 56, 417

\bibitem[{{Ul'yanov}, {Zakharenko} \& {Bruk}(2008){Ul'yanov}, {Zakharenko}, \&
  {Bruk}}]{2008ARep...52..917U}
{Ul'yanov} O.~M., {Zakharenko} V.~V., {Bruk} Y.~M., 2008, Astronomy Reports,
  52, 917

\bibitem[{{Volokitin}, {Krasnoselskikh} \& {Machabeli}(1985){Volokitin},
  {Krasnoselskikh}, \& {Machabeli}}]{volokitin1985waves}
{Volokitin} A.~S., {Krasnoselskikh} V.~V., {Machabeli} G.~Z., 1985, Fizika
  Plazmy, 11, 531

\bibitem[{{Zakharenko} {et~al}\mbox{.}(2007){Zakharenko}, {Nikolaenko},
  {Ulyanov}, \& {Motiyenko}}]{2007RRPRA..12..233Z}
{Zakharenko} V.~V., {Nikolaenko} V.~S., {Ulyanov} O.~M., {Motiyenko} R.~A.,
  2007, Radio Physics and Radio Astronomy, 12, 233

\bibitem[{{Zakharenko}, {Sharykin} \& {Rudavin}(2005){Zakharenko}, {Sharykin},
  \& {Rudavin}}]{2005KFNTS...5...90Z}
{Zakharenko} V.~V., {Sharykin} N.~K., {Rudavin} E.~R., 2005, Kinematika i
  Fizika Nebesnykh Tel Supplement, 5, 90

\bibitem[{{Zakharenko} {et~al}\mbox{.}(2013){Zakharenko}, {Vasylieva},
  {Konovalenko}, {Ulyanov}, {Serylak}, {Zarka}, {Grie{\ss}meier}, {Cognard}, \&
  {Nikolaenko}}]{2013MNRAS.431.3624Z}
{Zakharenko} V.~V. {et~al.}, 2013, MNRAS, 431, 3624

\bibitem[{{Zanimonskiy} {et~al}\mbox{.}(2010){Zanimonskiy}, {Zalizovski},
  {Lisachenko}, {Sopin}, \& {Yampolski}}]{2010RPRA..Zanimovskiy}
{Zanimonskiy} Y.~M., {Zalizovski} A.~V., {Lisachenko} V.~N., {Sopin} A.~O.,
  {Yampolski} Y.~M., 2010, Radio Physics and Radio Astronomy, 15, 164, in
  russian

\bibitem[{{Zheleznyakov}(1977)}]{1977ewcp.book.....Z}
{Zheleznyakov} V.~V., 1977, {Electromagnetic waves in cosmic plasma. Generation
  and propagation}. Nauka, Moscow, p. 432

\bibitem[{{Zheleznyakov}(1997)}]{Zheleznyakov1997}
{Zheleznyakov} V.~V., 1997, {Radiation in astrophysical plasmas}. Yanus-K,
  Moscow, p. 528

\end{thebibliography}
	
	\label{lastpage}
	
\end{document}